\input psfig.sty

\def\ptitle{A simple interpolation formula for the spectra of power-law and log potentials}
\nopagenumbers
\hsize 6.0 true in 
\hoffset 0.25 true in 
\emergencystretch=0.6 in                 
\vfuzz 0.4 in                            
\hfuzz  0.4 in                           
\vglue 0.1true in
\mathsurround=2pt                        
\topskip=20pt                            
\def\nl{\noindent}                       
\def\np{\hfil\vfil\break}                
\def\title#1{\bigskip\noindent\bf #1 ~ \trr\smallskip} 
\font\trr=cmr12                         
\font\bf=cmbx12                         
\font\sl=cmsl12                         
\font\it=cmti12                         
\font\trbig=cmbx12 scaled 1500          
\font\tiny=cmr10                         
\def\ng{>\kern -9pt|\kern 9pt}          
\def\hi#1#2{$#1$\kern -2pt-#2}          
\def\hy#1#2{#1-\kern -2pt$#2$}          

\def\sgn{\rm sgn}
\def\half{{1 \over 2}}


\output={\shipout\vbox{\makeheadline
                                      \ifnum\the\pageno>1 {\hrule}  \fi 
                                      {\pagebody}   
                                      \makefootline}
                   \advancepageno}

\headline{\noindent {\ifnum\the\pageno>1 
                                   {\tiny \ptitle\hfil page~\the\pageno}\fi}}
\footline{}
\newcount\zz  \zz=0  
\newcount\q   
\newcount\qq    \qq=0  

\def\pref #1#2#3#4#5{\frenchspacing \global \advance \q by 1     
    \edef#1{\the\q}
       {\ifnum \zz=1 { %
         \item{[\the\q]} 
         {#2} {\bf #3},{ #4.}{~#5}\medskip} \fi}}

\def\bref #1#2#3#4#5{\frenchspacing \global \advance \q by 1     
    \edef#1{\the\q}
    {\ifnum \zz=1 { %
       \item{[\the\q]} 
       {#2}, {\it #3} {(#4).}{~#5}\medskip} \fi}}

\def\gref #1#2{\frenchspacing \global \advance \q by 1  
    \edef#1{\the\q}
    {\ifnum \zz=1 { %
       \item{[\the\q]} 
       {#2}\medskip} \fi}}

 \def\sref #1{~[#1]}

\def\references#1{\zz=#1
   \parskip=2pt plus 1pt   
   {\ifnum \zz=1 {\noindent \bf References \medskip} \fi} \q=\qq
\pref{\quigg}{C. Quigg and J. L. Rosner, Phys. Lett.}{B 71}{153, 1977}{}
\pref{\mach}{M. Machacek and Y. Tomozawa, Ann. Phys. (N.Y.)}{110}{40, 1978}{}
\pref{\halld}{R. L. Hall and C. S. Kalman, Phys. Lett.}{B 83}{80, 1979}{}
\pref{\marta}{A. Martin, Phys. Lett.}{B 100}{511, 1981}{}
\pref{\martb}{A. Martin and J-M Richard, Phys. Lett.}{B 355}{345, 1995}{}
\pref{\quigga}{C. Quigg and J. Rosner, Phys. Rep }{C56}{167, 1979}{}
\pref{\gas}{S. Gasiorowicz and J. Rosner, Am. J. Phys.}{49}{954, 1981}{}
\pref{\hallb}{R. L. Hall, Phys. Rev.}{A 39}{5500, 1989}{} 
\pref{\halla}{R. L. Hall, J. Math. Phys.}{34}{2779 (1993)}{} 
\bref{\flug}{S. Fl\"ugge}{Practical Quantum Mechanics}{Springer, New York, 1974}{The linear potential is discussed on p101}

 }

\references{0}    

\trr 
\vskip 1.0true in
\centerline{\trbig A simple interpolation formula for the spectra}
\vskip 0.3true in
\centerline{\trbig of power-law and log potentials}
\vskip 0.5true in
\baselineskip 12 true pt 
\centerline{\bf Richard L. Hall}\medskip
\centerline{\sl Department of Mathematics and Statistics,}
\centerline{\sl Concordia University,}
\centerline{\sl 1455 de Maisonneuve Boulevard West,}
\centerline{\sl Montr\'eal, Qu\'ebec, Canada H3G 1M8.}
\vskip 0.2 true in
\centerline{email:\sl~~rhall@cicma.concordia.ca}
\bigskip\bigskip

\baselineskip = 18true pt  
\centerline{\bf Abstract}\medskip
Non-relativistic potential models are considered of the pure power $V(r) = \sgn(q) r^{q}$ and logarithmic $V(r) = \ln(r)$ types.  It is shown that, from the spectral viewpoint, these potentials are actually in a single family. The log spectra can be obtained from the power spectra by the limit $q\rightarrow 0$ taken in a smooth representation $P_{n\ell}(q)$ for the eigenvalues $E_{n\ell}(q).$  A simple approximation formula is developed which yields the first thirty eigenvalues with error $< 0.04\%.$  
\medskip\noindent PACS~~12.39.Pn;~12.39.Jh;~03.65.Ge.

\np
  \title{1.~~Introduction}
We consider a single particle that moves in a central potential $V(r)$ and obeys non-relativistic quantum mechanics.  We study two cases: (a) $V(r) = \sgn(q)r^{q},$ and (b) $V(r) = \ln(r).$  The main purpose of the paper is to show that these two problems are intimately related, and to provide a single formula which yields accurate approximations for the corresponding discrete Schr\"odinger eigenvalues.  The power-law and logarithmic potentials continue to be employed as non-relativistic models for quark confinement\sref{\quigg-\martb}. Hence it is important to understand that they are, in a sense, from the same family of potentials. A detailed analysis of the properties of non-relativistic potential models is given in the review article by Quigg and Rosner\sref{\quigga}. The approach to this question based on the expression $\{dr^{q}/dq\}_{q\rightarrow 0} = \ln r$ is also been discussed in Ref.[\gas]. In the present paper we approach the same problem by using what we call the `$P$-representation' for the discrete spectrum generated by power-law potentials. This allows us to construct a general energy eigenvalue formula which joins $q = -1$ smoothly to $q = 2,$ passing through the logarithmic case $q = 0.$\medskip

The power-law family and logarithmic potential will be kept distinct until it becomes clear in what sense the logarithmic potential corresponds to the limit $q \rightarrow 0$ of the power-law class.  We first treat the elementary issue of scaling.  We write the eigenvalues of the bare problems as:

$$\cases{
-\Delta + \sgn (q)r^{q} & $\longrightarrow E_{n\ell}(q)$\cr
&\cr
-\Delta + \ln (r)       & $\longrightarrow E_{n\ell}^{L},$\cr
}\eqno{(1.1)}$$

\nl where $n = 1,2,3,\dots$ counts the discrete eigenvalues in each angular-momentum subspace labelled by $\ell = 0,1,2,\dots.$ The eigenvalues so labelled have degeneracy precisely $2\ell+1.$  Scaling arguments\sref{\hallb,\halld} can be used to show that more general Hamiltonians have the corresponding eigenvlues given in terms of $\{E_{n\ell}(q),~E_{n\ell}^{L}\}$ by

$$\cases{
-\mu\Delta + v\ \sgn (q)r^{q} & $\longrightarrow \mu\left ({v\over\mu}\right)^{2\over{2+q}}E_{n\ell}(q)$\cr
&\cr
-\mu\Delta + v\ln (r)       & $\longrightarrow vE_{n\ell}^{L} - \half v\ln\left({v\over\mu}\right),$\cr
}\eqno{(1.2)}$$

\nl where $v > 0$ is a coupling parameter and $\mu > 0$ might, for example, be $\mu = \hbar^{2}/(2m).$  As a consequence of these scaling rules we shall only need to consider the special case $\mu = v = 1$ for the rest of the paper.\medskip

In Section~(2) we introduce an exact semi-classical representation $P_{n\ell}(q)$ for these eigenvalues which will allow us to show how the log spectrum can be derived as a limit from the family of power spectra.  As a motivation for this, we plot the power eigenvalues $E_{n\ell}(q)$ as functions of $q$ in Fig.(1).  These graphs are to be compared with the corresponding smooth $P$-representation shown in Fig.(2).  Once the $P$-representation is established, we go on to consider the {\it log-power theorem} $P_{n\ell}(0) = P_{n\ell}^{L},$ and, in Section~(3), to develop an approximation formula for $P_{n\ell}(q).$  
  
  \title{2.~~The functions $E(q)$ and $P(q)$}
The $P$-representation for the spectrum is as follows:
$$E_{n\ell} = \min_{r > 0}\left\{{{P_{n\ell}^2}\over {r^2}} + V(r)\right\}.\eqno{(2.1)}$$
This transformation is well defined provided we can always find the $P$ which corresponds to a given $E.$  For smooth potentials, the minimum exists and is unique if $r^{3}V'(r)$ is monotone, a condition which is certainly satisfied if $V(r)$ is a pure power or log.  For these potentials, Eq.(2.1) is therefore an exact semi-classical representation for the eigenvalues: the kinetic energy term scales like $L^{-2},$ as it should. It is perhaps interesting also to note that positive factors such as $\mu$ and $v$ can be re-introduced in front of the kinetic- and potential-energy terms, without having to revise the $P$ numbers.  For the specific pure-power and log problems at hand, we have respectively:
$$\eqalign{E_{n\ell}(q) &= \min_{r > 0}\left\{{{P_{n\ell}^2}(q)\over {r^2}} + \sgn(q)r^{q}\right\}\cr
&= \sgn(q)\left({q\over 2} + 1\right)\left({{2P_{n\ell}^{2}}\over {|q|}}\right)^{q\over {q+2}}\cr}\eqno{(2.2)}$$
and
$$\eqalign{E_{n\ell}^{L} &= \min_{r > 0}\left\{{{P_{n\ell}^L}^{2}\over {r^2}} + \ln(r)\right\}\cr
&= \ln\left((2e)^{\half} P_{n\ell}^{L}\right).\cr}\eqno{(2.3)}$$

\nl Some of the $E,$ and the corresponding $P,$ are already known exactly, from elementary quantum mechanics. Thus from the known eigenvalues for the Hydrogen atom $E_{n\ell}(-1) = -[2(n+\ell)]^{-2}$ and the harmonic oscillator $E_{n\ell}(2) = 4n + 2\ell -1$ we immediately obtain the outer $P$ numbers of our range of interest $-1 \leq q \leq 2:$

$$P_{n\ell}(-1) = n+\ell,\quad P_{n\ell}(2) = 2n + \ell - \half.\eqno{(2.4)}$$

\nl Our goal is now to interpolate $P_{n\ell}(q)$ at interior points of the range. 

The spectral relationship of pairs of potentials $\{U,\ V\}$ that are smooth transformations $V(r) = g(U(r))$ of one another may be analysed according to the following geometrical reasoning\sref{\halla}. We suppose that the eigenvalues of $H^{[U]} = -\Delta + v U(r)$ are given by $E_{n\ell}^{[U]}(v).$ Each tangent to $V = g(U)$ is a potential of the form $V^{t}(r) = \alpha(t)U(r) + \beta(t),$ where $t$ is the point of contact, and $\alpha(t) = g'(U(t))$ and $\beta(t) = g(U(t)) - U(t)g'(U(t)).$  When $g$ has definite convexity, every tangential potential lies on one side of $V$ and we obtain energy bounds; an optimization over $t$ yields the best such bound.  In the concave case ($g'' < 0$), for example, we obtain:
$$E^{[V]}_{n\ell}(v) < \min_{t > 0}\left\{E_{n\ell}^{[U]}(v\alpha(t)) + \beta(t)\right\}.\eqno{(2.5)}$$
\noindent The analytical complexity of the optimization in (2.5) is greatly reduced by the introduction of `kinetic potentials'\sref{\halla}, which, in the case of power-law potentials, lead, after a change of variable and the correct choice of $P,$ to the very simple form (2.1). It is straightforward to show for the present problem that each power potential with power $q_1$ is a convex transformation of all other power potentials with powers $q_2 < q_1.$  Meanwhile, the log potential is a convex function of each negative power, and, at the same time, it is a concave transformation of each positive power.  Two results which follow from this analysis are (i)\sref{\hallb} the functions $P_{n\ell}(q)$ are monotone increasing in $q,$  and (ii)\sref{\halla} the {\it log-power theorem}, which says that
$\lim_{q \rightarrow 0}P_{n\ell}(q) = P_{n\ell}(0) = P_{n\ell}^{L}.$ 

As is clear from Fig.(1), the spectra $E_{n\ell}(q)$ of the power potentials converge on the values $\pm 1$ as the power $q$ approaches zero respectively from positive or negative values. Meanwhile, the set of points for the log spectrum $E_{n\ell}^{L}$ would appear to be quite unrelated, and would have no natural place on this graph.  In the $P$-representation the picture is quite different, as is shown in Fig.(2).  The monotone $P(q)$ curves are distinct at $q = 0,$ and we know that they are exactly equal there to the values corresponding to the log potential.  This smooth expression of the entire spectral family suggests that we can interpolate the $P$ curves with a simple polynomial, a task which is taken up in the next Section. We made an earlier attempt at an interpolation\sref{\hallb}, which was much less accurate because it could not benefit from the (yet undiscovered) log-power theorem. 
  \title{3.~~Interpolation formula for $P(q)$}
As a model for the $P$ curves we use a cubic polynomial which allows us to fit values at the four points $q = \{-1,0,1,2\};$ the exact formulas (2.4) are used for $q = \{-1, 2\}.$ We make the expansion about $q = 0$ so as to favour the $q$ values between log and linear that are used in potential models.  Thus we define
$$P(q) = a + b q + c q^{2} + d q^{3},\eqno{(3.1)}$$
\nl in which, for simplicity, we have omitted the quantum-number subscripts $n\ell.$  By inversion we have
$$\left[\matrix{a\cr b\cr c\cr d}\right] = \left[
\matrix{
0&1&0&0\cr
-{1\over 3} &-{1\over 2}&1&-{1\over 6}\cr
\half &-1&\half &0\cr
-{1\over 6}&\half&-\half&{1\over 6}
}\right]\left[\matrix{P(-1)\cr P(0)\cr P(1)\cr P(2)}\right]\eqno{(3.2)}.$$
\nl For the linear potential $q = 1$ the exact S-state eigenvalues may be expressed in terms of the zeros of the Airy function\sref{\flug}. We take these known values and complement them with others computed numerically for the log and linear potentials.  By inverting Eq.(2.2) and Eq.(2.3) we have thus computed the coefficients $P(0) = {\rm log}$ and $P(1).$  The $P$-data for the first 30 eigenvalues are exhibited in Table(1), along with the approximations we get for the energies $E_{n\ell}(\half)$   For comparison, we have tabulated also the percentage errors in these eigenvalue results.  For the first 30 eigenvalues, the percentage errors are all positive and less than $0.04\%$ at $q = \half;$  as $q$ approaches $q = 0$ or $q = 1,$ these errors decrease dramatically.  Similar accuracy is obtained for the whole range $-1 \leq q \leq 2$ of the interpolation.
  \title{4.~~Conclusion}
In spite of present-day computing convenience, it is still very useful to have an approximate eigenvalue formula, particularly a simple and accurate one.  The possibility of a simple formula is a consequence of the existence of the smooth monotone $P$-representation for the eigenvalues.   Since the early work on potential models there has been a realization that, from a practical point of view, a log and a power potential, with $0 < q < 1,$ could be adjusted to serve in the model almost equally well, especially if $q$ is small.  The log-power theorem provides a theoretical basis for these concrete spectral observations.  
   \title{Acknowledgment}
Partial financial support of this work under Grant No. GP3438 from the Natural Sciences and Engineering Research Council of Canada is gratefully acknowledged. 
\np
 \references{1}
\np
\noindent {\bf Table 1}~~The `input' values of $P_{n\ell}(0)$ and $P_{n\ell}(1);$ and the approximations $E^{A}_{n\ell}(\half)$ for $E_{n\ell}(\half)$ obtained via the cubic $P$ formula (3.1), with the percentage errors.
\baselineskip=16 true pt 
\def\vr{\vrule height 12 true pt depth 6 true pt}
\def\vra{\vr\hfill} \def\vrb{\hfill &\vra} \def\vrc{\hfill & \vr\cr\hrule}
\def\vrq{\vr\quad} 

$$\vbox{\offinterlineskip
 \hrule
\settabs
\+ \vrq \kern 0.4true in &\vrq \kern 0.4true in &\vrq \kern 0.7true in &\vrq \kern 0.7true in &\vrq \kern 0.7true in &\vrq \kern 0.7true in &\vrq \kern 0.5true in &\vr\cr\hrule
\+ \vra $n$ \vrb $\ell$\vrb $P_{n\ell}(0)$\vrb $P_{n\ell}(1)$\vrb $E^{A}_{n\ell}(\half)$\vrb $E_{n\ell}(\half)$\vrb \% \vrc
\+ \vra 1\vrb 0\vrb 1.21867\vrb 1.37608\vrb 1.83375\vrb 1.83339\vrb 0.019\vrc
\+ \vra 2\vrb 0\vrb 2.72065\vrb 3.18131\vrb 2.55142\vrb 2.55065\vrb 0.030\vrc
\+ \vra 3\vrb 0\vrb 4.23356\vrb 4.99255\vrb 3.05224\vrb 3.05118\vrb 0.035\vrc
\+ \vra 4\vrb 0\vrb 5.74962\vrb 6.80514\vrb 3.45341\vrb 3.45213\vrb 0.037\vrc
\+ \vra 5\vrb 0\vrb 7.26708\vrb 8.61823\vrb 3.79482\vrb 3.79336\vrb 0.039\vrc
\+ \vra 1\vrb 1\vrb 2.21348\vrb 2.37192\vrb 2.30073\vrb 2.30050\vrb 0.010\vrc
\+ \vra 2\vrb 1\vrb 3.68538\vrb 4.15501\vrb 2.85486\vrb 2.85434\vrb 0.018\vrc
\+ \vra 3\vrb 1\vrb 5.17774\vrb 5.95300\vrb 3.28659\vrb 3.28583\vrb 0.035\vrc
\+ \vra 4\vrb 1\vrb 6.67936\vrb 7.75701\vrb 3.64835\vrb 3.64739\vrb 0.026\vrc
\+ \vra 5\vrb 1\vrb 8.18607\vrb 9.56408\vrb 3.96382\vrb 3.96268\vrb 0.029\vrc
\+ \vra 1\vrb 2\vrb 3.21149\vrb 3.37018\vrb 2.65775\vrb 2.65756\vrb 0.007\vrc
\+ \vra 2\vrb 2\vrb 4.66860\vrb 5.14135\vrb 3.12077\vrb 3.12033\vrb 0.014\vrc
\+ \vra 3\vrb 2\vrb 6.14672\vrb 6.92911\vrb 3.50309\vrb 3.50245\vrb 0.018\vrc
\+ \vra 4\vrb 2\vrb 7.63639\vrb 8.72515\vrb 3.83336\vrb 3.83254\vrb 0.021\vrc
\+ \vra 5\vrb 2\vrb 9.13319\vrb 10.52596\vrb 4.12678\vrb 4.12581\vrb 0.024\vrc
\+ \vra 1\vrb 3\vrb 4.21044\vrb 4.36923\vrb 2.95461\vrb 2.95445\vrb 0.005\vrc
\+ \vra 2\vrb 3\vrb 5.65879\vrb 6.13298\vrb 3.35798\vrb 3.35759\vrb 0.012\vrc
\+ \vra 3\vrb 3\vrb 7.12686\vrb 7.91304\vrb 3.70327\vrb 3.70270\vrb 0.015\vrc
\+ \vra 4\vrb 3\vrb 8.60714\vrb 9.70236\vrb 4.00810\vrb 4.00737\vrb 0.018\vrc
\+ \vra 5\vrb 3\vrb 10.09555\vrb 11.49748\vrb 4.28283\vrb 4.28196\vrb 0.020\vrc
\+ \vra 1\vrb 4\vrb 5.20980\vrb 5.36863\vrb 3.21247\vrb 3.21233\vrb 0.004\vrc
\+ \vra 2\vrb 4\vrb 6.65235\vrb 7.12732\vrb 3.57310\vrb 3.57275\vrb 0.010\vrc
\+ \vra 3\vrb 4\vrb 8.11305\vrb 8.90148\vrb 3.88950\vrb 3.88898\vrb 0.013\vrc
\+ \vra 4\vrb 4\vrb 9.58587\vrb 10.68521\vrb 4.17335\vrb 4.17268\vrb 0.016\vrc
\+ \vra 5\vrb 4\vrb 11.06163\vrb 12.47532\vrb 4.43164\vrb 4.43131\vrb 0.008\vrc
}$$
\np
\baselineskip 18 true pt 

\hbox{\vbox{\psfig{figure=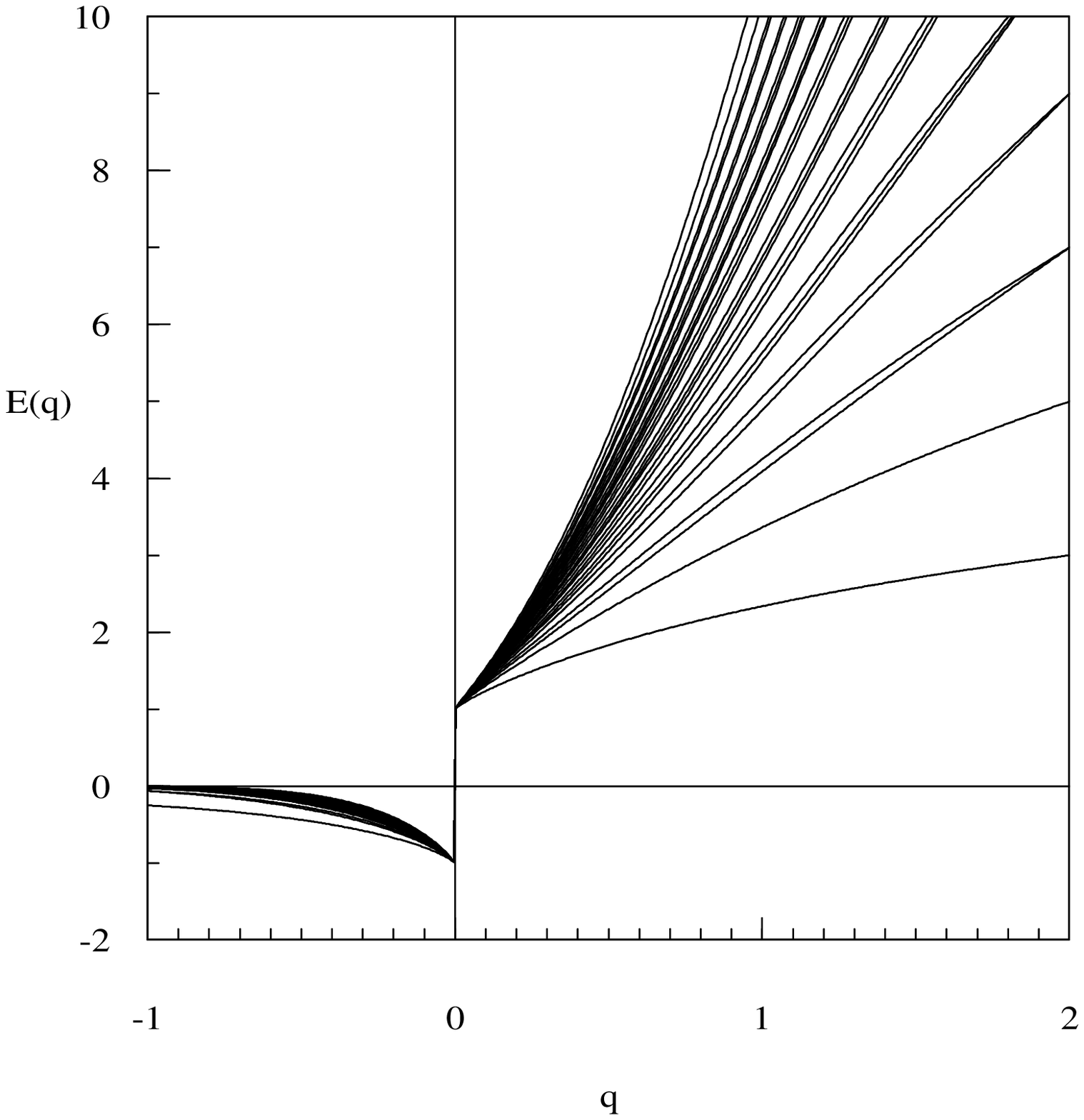,height=6in,width=5in,silent=}}}

\title{Figure 1.}
\nl The first 30 eigenvalues $E_{n\ell}(q),$ $1 \leq n \leq 5,$ $0 \leq \ell \leq 5,$ corresponding to the power potential $V(r) = \sgn(q) r^{q}.$  For $q > 0,$ the eigenvalues increase with $q$  from $1$ to $E_{n\ell}(2) = 4n + 2\ell + 1;$~  for $q < 0,$ they decrease (as $q$ increases) from $E_{n\ell}(-1) = -[2(n+\ell)]^{-2}.$ to $-1.$  Both sets of curves increase with $n$ and $\ell.$
\hfil\vfil\break
\hbox{\vbox{\psfig{figure=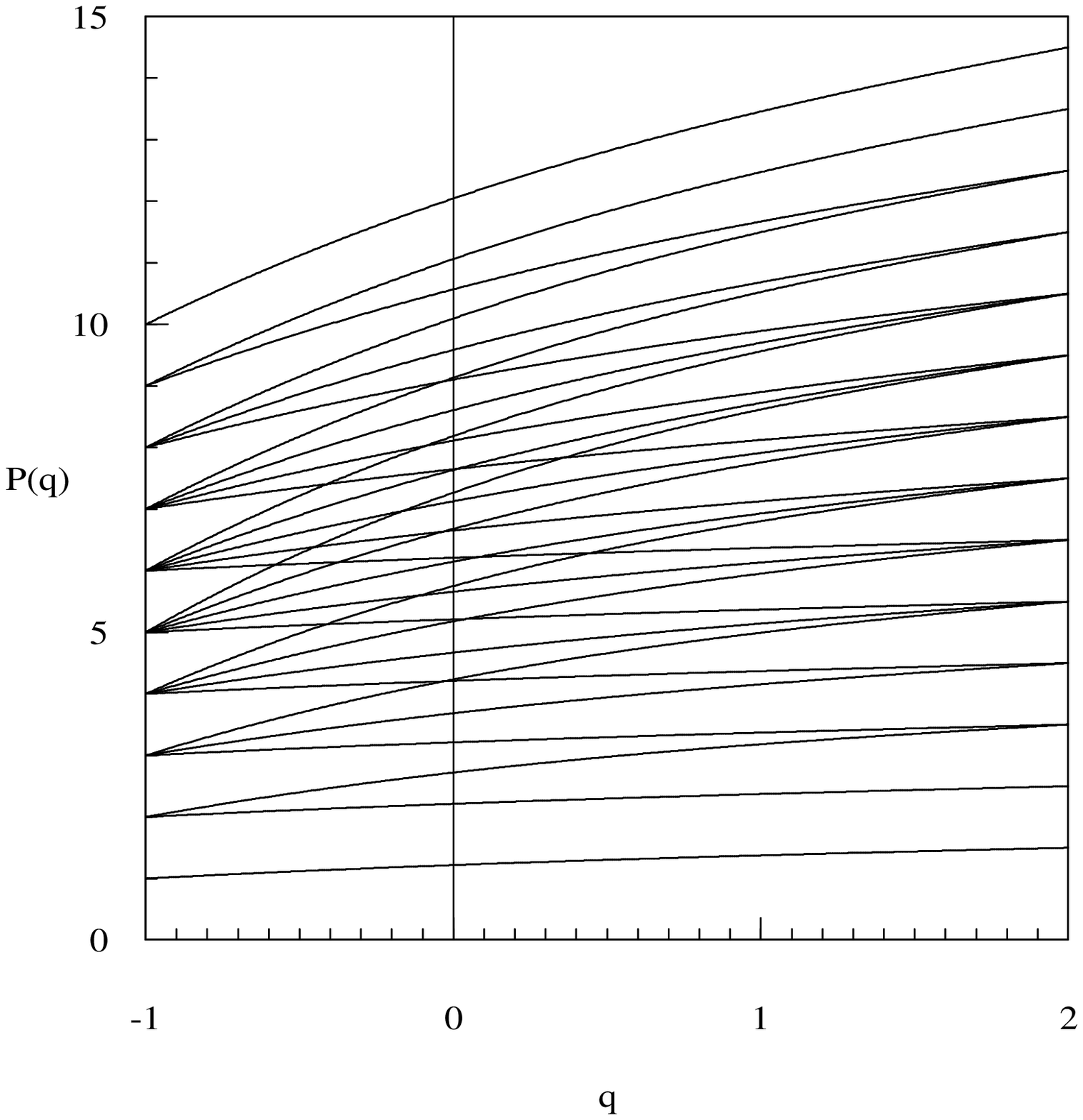,height=6in,width=5in,silent=}}}

\title{Figure 2.}
\nl In the $P$-representation, the same set of 30 eigenvalues shown in Fig.(1) now lie on monotone smooth curves.  The log-power theorem states that the $P$ values for the log potential are precisely $P_{n\ell}(0).$ As $q$ increases from $-1$ to $2$, the degeneracy of the Coulomb problem $P_{n\ell}(-1) = n + \ell$ evolves into the degeneracy of the harmonic oscillator $P_{n\ell}(2) = 2n + \ell - \half.$
\end